\documentstyle[12pt]{article}
\newcommand{\be}{\begin{equation}}
\newcommand{\ee}{\end{equation}}
\newcommand{\ed}{\end{document}}
\newcommand{\lab}[1]{\label{#1}}
\newcommand{\re}[1]{(\ref{#1})}
\newcommand{\ci}[1]{\cite{#1}}
\renewcommand{\baselinestretch}{1.4}
\date{}
\title{CHAOTIC AUTOIONIZATION OF RELATIVISTIC TWO-ELECTRON ATOM}
\author{ D.U.MATRASULOV \\
Heat Physics Department of the Uzbek Academy of Sciences,\\
28 Katartal St.,700135 Tashkent, Uzbekistan }
\begin{document}\large
\maketitle

\begin{abstract}
Chaotic autoionization of relativistic two-electron
atom in the monochromatic field is investigated. A theoretical analysis
of chaotic dynamics of the relativistic outer electron under the periodic
perturbation due to the inner electron, based on Chirikov
criterion is given. The diffusion
coefficient, the ionization rate and time are calculated.
\end{abstract}

Study of highly excited one-and two-electron atoms on the basis of classical
mechanics has been the subject of extensive theoretical \ci{del83,jens84,cas88}
as well as experimental \ci{bay74,jens91} investigation recently.
One of the most interesting phenomena appearing in the highly excited atom interacting
with monochromatic field is the chaotization of motion of the classical electron,
that leads to diffusive (or chaotic) ionization.
Application of
methods of stochastic dynamics to such interaction gives an important
tool for these studies.
In particular,
the investigation of chaotization of motion of Kepler electron under the influence
of monochromatic field using the Chirikov criterion allows one to estimate critical
value of the field strength, at which diffusive ionization will occur.
The same type of ionization to be suggested in the two-electron atom even in the
absence of external monocharomatic field.  The motion of the outer electron can
become chaotic due to the interaction with the periodically moving inner electron
that leads to the chaotic autoionization of the outer electron. Thus the mechanism
of classical autoionization is the following: under the influence of periodic
perturbation due to the inner electron the outer one passes to more excited
orbits and escape to infinity. Such a mechanism was first investigation
by Krainov and Belov\ci{bel87}. In this Brief Report we generalize their results to
the relativistic case. One should note that up to now investigations of deterministic
chaos were mainly limited by the consideration of nonrelativistic systems, though
there are few works in which chaotic dynamics of relativistic systems
are considered \ci{cher89,luch96,kim96}. In our recent paper\ci{mat} we have generalized
chaotic ionization of hydrogen atom by monochromatic field to the case of
relativistic hydrogenlike atom. Further everywhere we use the relativistic
system of units $m_{e}=\hbar=c=1$.

Before considering the
classical chaotic dynamics of the relativistic two-electron atom
 we give brief classical description of the relativistic Kepler motion in terms of action-angle
variables. The Hamiltonian of the relativistic Kepler electron in the action-angle
variables is \ci{born}
\be
H_{0} = [1+ \frac{Z^2\alpha^2}{(n-M+\sqrt{M^2 -Z^2\alpha^2})^2}]^{-\frac{1}{2}},
\lab{hamilt1}
\ee
where $n = I_{r}+I_{\phi}+I_{\theta}, \;\;\;$ $M=I_{\phi}+I_{\theta},$

$I_{r},I_{\phi},I_{\theta}$ are the radial and angular components of the action.

Trajectory equation of relativistic electron is given by \ci{land}
\be
\frac{p}{r} +ecosq\Phi-1,
\lab{traject}
\ee
where
\be p= \frac{M^2-Z^2\alpha^2}{EZ\alpha}, \;\; q=\sqrt{1-\frac{Z^2\alpha^2}{M^2}},\;\;
e= \sqrt{1-\frac{M^2-Z^2\alpha^2}{n^2}},
\lab{exc}
\ee
$E$ is the energy of the electron.

Due to the factor $q$ trajectory of the relativistic Kepler electron  is not
closed \ci{land}.
For
\be
M\gg Z\alpha = \frac{Z}{137}
\lab{cond1}
\ee
the Hamiltonian \re{hamilt1} takes the form
\be
H_{0} = [1+ \frac{Z^2\alpha^2}{(n-M+M)^2}]^{-\frac{1}{2}}\approx ,
\frac{n}{\sqrt{n^2 +Z^2\alpha^2}}.
\ee
In this approximation $q\approx 1$ and trajectory becomes closed. Radius $r$
and polar angle $\phi$ in terms of action-angle variables can be written as
\be
r = \frac{n\sqrt{n^2 +Z^2\alpha^2}}{Z}(1-ecos\psi),
\lab{radius}
\ee
\be
ctg\phi = \frac{n}{\gamma}tg\frac{\psi}{2},
\lab{angle}
\ee
where $\gamma = \sqrt{M^2 - Z^2\alpha^2}$.

Consider now a relativistic two-electron atom. For the simplicity we will
assume that both electrons move on the same plane. Electrons interact with the
atomic core of charge $Z\alpha$. The Hamiltonian of two-electron atom can
be written as
\be
T_{1}+-\frac{Z}{r_{1}}+ T_{2} -\frac{Z}{r_{2}} +\frac{1}{\mid{\bf r_1}-{\bf r_2}\mid},
\lab{hamilt4}
\ee
where $T_{i}$ is the kinetic energies of the relativistic electrons, $r_1$ and $r_2$ are
the distances from the atomic core to the inner and outer electrons, respectively;
the last term in this expression describes the interelectronic repulsion. Assuming
$r_1\ll r_2$ the last term in \re{hamilt4} can be written as
\be
\frac{1}{\mid {\bf r_1}-{\bf r_2}\mid} \approx \frac{1}{r_1}- \frac{{\bf r_1 r_2}}{r_2^3}.
\lab{dipole}
\ee
Inserting \re{dipole} into \re{hamilt4} we have
\be
T_{1}+-\frac{Z}{r_{1}}+ T_{2} +\frac{Z'}{r_{2}} +V,
\lab{hamilt5}
\ee
where
$$
V= \frac{{\bf r_1 r_2}}{r_2^3}\cos(\phi_1-\phi_2),
$$
$Z'=Z-1\;\;$, $\phi_1\;\;$,$\phi_2$ are the azimutal angles of the electrons on
the plane of motion.

To simplify calculations we assume that the inner electron moves along the circular
orbit, i.e. without eccentricity. Then from \re{radius}  we have
\be
r_1 = \frac{n_1\sqrt{n_{1}^2 +Z^2\alpha^2}}{Z}
\lab{radius1}
\ee
\be
\phi = \omega_1 t = \frac{Z^2\alpha^2}{(n_1^2+Z^2)^\frac{2}{3}}
\lab{angle1}
\ee

Assuming, as in \ci{bel87} $M_2/I_2\ll1,\;\;$ $cos\psi\approx 1$, one can obtain
for the outer electron
$$
r_2 = \frac{2n_2\sqrt{n_2^2 +Z'^2\alpha^2}}{Z'}(sin^2\frac{\psi}{2}+\frac{a^2}{4}),
$$
where $a = \gamma^2/n_2^2$.

Thus perturbation can be written as
$$
V = \frac{Z'^2\alpha^2}{4n_2^2(n_2^2+Z^2\alpha^2)}r_1(sin^2\frac{\psi}{2}+\frac{a^2}{4})^{-2}
cos\omega t
$$

Using the Fourier expansion for the dipole moment from \ci{bel87}
$$
d = (sin^2\frac{\psi}{2}+\frac{a^2}{4})^{-2} = \sum d_k cos(k\theta_2),
$$

where $d_k = 26.3/(a^6 k^{\frac{5}{3}})$,
the full Hamiltonian can be written as
\be
H =\frac{n_2}{\sqrt{n_2^2 +Z'^2\alpha^2}}+\epsilon cos \omega_1 t\sum d_k cos(k\theta_2),
\ee
with
$$
\epsilon = \frac{Z'^2\alpha^2}{4n_2^2(n_2^2+Z^2\alpha^2)}r_1
$$

This Hamiltonian is equivalent to the one describing the interaction of relativistic
hydrogenlike atom  with monochromatic field \ci{mat}(though here $d_k$ and $\epsilon$
are defined by another formulae). Therefore chaotic dynamics of the
outer electron can be treated by the same method as in \ci{mat}. For the resonance
width we have \ci{mat}
\be
\Delta\omega = 4(2\frac{\omega}{dn}r_{1}r_{2}d_k)^{\frac{1}{2}}
\lab{Delta}
\ee
The distance between two neighboring resonances is
\be
\delta\omega = \frac{\omega_1}{k^2}
\lab{delta}
\ee
The Chirikov criterion for resonance overlap can be written as \ci{bel87}
$$
2.5\frac{\Delta\omega}{\delta\omega}>1,
$$
This gives us (using \re{Delta} and \re{delta})
\be
400Z'^{-\frac{2}{3}}Z^{-\frac{1}{3}}n_1 n_2^5 \gamma^{-6}>1
\ee
Analogically to that as was done in \ci{bel87,mat}  one can calculate diffusion coefficient
$$
D = \frac{\pi}{2}\frac{r_1^2r_2^{-4}k^2d_{k}^2}{\omega_{2}^2}
$$
Expressing $r_1,\;r_2,\; d_{k}\;$ and $\omega_{2}$ via actions and charges one can obtain
$$
D \approx 68 Z'^{\frac{8}{3}}Z^{-\frac{14}{3}}(M_2^2-Z^2\alpha^2)^{-6}
n_1^2(n_1^2+Z^2\alpha^2)^3n_2^8(n_2^2+Z'^2\alpha^2)^{-1}
$$
This diffusion coefficient can be written in terms of nonrelativistic one and
first-order relativistic corrections as following:

\be
D \approx D_{nonrel}(1+3\frac{Z^2\alpha^2}{n_1^2})(1+6\frac{Z^2\alpha^2}{M_2^2})
(1-\frac{Z'^2\alpha^2}{n_2^2})
\lab{reldif}
\ee
where
$$
D_{nonrel} \approx 68 Z'^{\frac{8}{3}}Z^{-\frac{14}{3}}n_1^8n_2^6/M_2^{12}
$$
is the nonrelativistic diffusion coefficient.

As is seen from \re{reldif} relativistic corrections to the diffusion coefficient
come due to the three factors:
first correction comes from increasing of the frequency of motion of
the inner electron and can be considerable; second one is due to the ratio $Z\alpha/M_{2}$ and
can be also considerable for small orbital moments; third one comes due to the
relativistic motion of inner electron and leads to slight decreasing of diffusion
coefficient.

Using \re{reldif} one can also calculate  the ionization time \ci{bel87}
$$
\tau_{D} \approx \frac{n_2^2}{2D}\approx
\tau_{nonrel}(1-3\frac{Z^2\alpha^2}{n_1^2})(1-6\frac{Z^2\alpha^2}{M_2^2})
(1+\frac{Z'^2\alpha^2}{n_2^2}).
$$
Then the ionization rate per unit of time can be written as
$$
\Gamma = \frac{2D}{n_2^2} \approx
\Gamma_{nonrel}(1+3\frac{Z^2\alpha^2}{n_1^2})(1+6\frac{Z^2\alpha^2}{M_2^2})
(1-\frac{Z'^2\alpha^2}{n_2^2})
$$

The above formulae show that the relativistic corrections to the diffusion coefficient,
autoionization time and rate are appreciable. One should note that the above results are valid
for the case when $M\gg Z\alpha$. For $M > Z\alpha$  the Hamiltonian \re{hamilt1} becomes
complex and the finite sizes of nucleus should be accounted. More detail analysis
of the considered above problem should be given by solving the classical equations of
motion.

The author thanks Prof. V.I.Matveev for helpful discussions.

\ed